\documentclass[5p,10pt]{elsarticle}
\usepackage[english]{babel}
\usepackage{amsmath,amssymb,amsthm,mathtools}
\usepackage{euscript,mathrsfs}
\usepackage{enumitem}
\usepackage{graphicx}
\usepackage{wrapfig}
\usepackage[dvipsnames]{xcolor}
\usepackage{indentfirst}
\usepackage{amsbsy}
\usepackage{wasysym}
\usepackage{cancel}
\usepackage{verbatim}
\usepackage{empheq}
\usepackage{adjustbox}
\usepackage[pdfencoding=auto]{hyperref}
\definecolor{urlcolor}{HTML}{990000}
\definecolor{linkcolor}{HTML}{005F5F}
\definecolor{refcolor}{HTML}{298C2C}
\hypersetup{pdfstartview=FitH,  linkcolor=refcolor,urlcolor=linkcolor, colorlinks=true,citecolor=blue}

\renewcommand{\phi}{\varphi}

\DeclareMathOperator{\sgn}{\mathrm{sgn}\,}

\begin{document}

\title{On out-of-equilibrium phenomena in pseudogap phase of complex SYK+U model}

\author[add1,add2]{Artem Alexandrov}
\ead{aleksandrov.aa@phystech.edu}
\address[add1]{Moscow Institute of Physics and Technology, Dolgoprudny, 141700, Russia}
\address[add2]{Laboratory of Complex Networks, Center for Neurophysics and 
Neuromorphic Technologies, Moscow, Russia}

\author[add2,add3]{Alexander Gorsky}
\address[add3]{Institute for Information Transmission Problems, Moscow, 127994, Russia}

\begin{abstract}
    In this Letter we consider the out-of-equilibrium phenomena in the complex Sachdev-Ye-Kitaev (SYK) model supplemented with the attractive Hubbard interaction (SYK+U). This model provides the clear-cut transition from non-Fermi liquid phase in pure SYK to the superconducting phase through the pseudogap phase with non-synchronized Cooper pairs. We investigate the quench of the phase soft mode in this model and the relaxation to the equilibrium state. Using the relation with Hamiltonian mean field (HMF) model we show that the SYK+U model enjoys the several interesting phenomena, like existence of quasi-stationary long living states, out-of-equilibrium finite time phase transitions, non-extensivity and tower of condensates. We comment on the holographic dual gravity counterparts of these phenomena.
\end{abstract}

\maketitle

\section{Introduction}

Rigorous theoretical description of high-temperature (high-T) superconductivity is an ambitious \& challenging task in condensed matter. One of the important actors on the scene of high-T superconductivity is so-called pseudogap phase \cite{Franz2007}. Such phase is observed experimentally by different techniques in many high-T superconductors \cite{Timusk1999}. Despite many efforts in theoretical investigation of pseudogap phase, the comprehensive view is still absent. Moreover, there is no strict definition of this phase. For our best knowledge, one of the most conventional treatment of pseudogap relies upon a non-zero gap with fluctuating phase, so-called preformed pairs scenario \cite{Loktev2001}. There are several models that describe the appearance of a pseudogap phase \cite{Varma2006,Imada2013,Chowdhury2016,Zhang2020}. 

In this Letter we focus on recently proposed pseudogap phase in Sachdev-Ye-Kitaev (SYK) model \cite{sachdev1993gapless,kitaev2015} with additional Hubbard interaction, which we hereafter call complex SYK+U model. In the pure complex SYK model the effective action for the phase field has been found 
and discussed in details in \cite{davison2017thermoelectric,gu2020notes}.
In \cite{Gorsky2020} it was demonstrated that in complex SYK+U model in the limit of small Hubbard interaction, $U\ll J$ (here $U$ is the Hubbard interaction constant and $J$ is the SYK model interaction constant) the pseudogap phase appears. This phase corresponds to a non-trivial saddle-point in mean-field treatment, where the phases of gaps $\Delta_i=|\Delta|e^{i\theta_i}$ are not fixed by saddle point equation. These phases are soft degrees of freedom and their fluctuations can destroy off-diagonal long-range order (ODLRO) even in $N\rightarrow \infty$ limit. The pseudogap phase corresponds to the non-synchronized Cooper pairs.
 
We shall use in this study that dynamics of the phase mode in SYK+U model is identical to the Kuramoto model with inertia \cite{Olmi2014} or, more correctly, to the Hamiltonian mean field (HMF) model. The HMF model is the toy-model for investigation of systems with long-range interactions (LRI). The LRI are modeled by all-to-all coupling (which sometimes referred as infinite range couplings) and in classical case cause interesting and well-known phenomena like violent relaxation (VR) \cite{Lynden1967,Giachetti2019,Santini2022}, existence of quasi-stationary states (QSS) with large lifetime \cite{Antoni1995,Barre2002,Barre2002out,Barre2009,Yamaguchi2004,Antoniazzi2007,Jain2007,Spohn2012,Koyama2008}, ensemble inequivalence \cite{Dauxois2000} and non-extensivity \cite{Campa2009,Campa2014Book}. During time evolution, the system is trapped in these QSS that do not obey Boltzmann statistics and can be treated in terms of Lynden-Bell \& core-halo distributions \cite{Levin2014}. For the SYK+U model with $U\ll J$, we will show that such states exist in the pseudogap phase. It turns out that the model possesses the equilibrium quantum phase transition from pseudogap to superconductor, which can be treated as transition between non-synchronized and synchronized states in quantum HMF model. The existence of QSS in quantum HMF model means that even in pseudogap phase partial synchronization can take place. This behavior is the quantum analog of QSS states in classical HMF \cite{Plestid2018}.

The generalization of SYK+U model to SYK+U dots array has been formulated in \cite{chudnovskiy2022superconductor}. It was found there that the model enjoys several nontrivial phases with the different transport properties. Another way to perform the transition from SYK to superconductivity involves the Yukawa interaction \cite{patel2018coherent,wang2020solvable,esterlis2019cooper,hauck2020eliashberg,chowdhury2020intrinsic,chowdhury2020unreasonable}. In particular the holographic picture in SYK+Yukawa has been developed in \cite{inkof2022quantum}. However the SYK+Yukawa model is not suitable enough for the investigation of pseudogap phase.

We shall use the relation between quantum dynamics of phase mode in SYK+U model and quantum HMF model to make the predictions concerning the out-of-equilibrium dynamics of the pseudogap phase.

\section{HMF model in SYK+U model}

In this Section we briefly recall the derivation of HMF model in complex SYK+U model. The model involves the complex fermions with four-fermion interaction and random coupling $J_{ijkl}$,
\begin{multline}
    H_{\text{SYK}} = \\ = \frac{1}{2} \sum_{ijkl,\sigma,\sigma'} J_{ijkl} \left[c^{\dagger}_{i \sigma}c^{\dagger}_{j \sigma'}c_{k \sigma'} c_{l \sigma} +  c^{\dagger}_{l \sigma}c^{\dagger}_{k \sigma'}c_{j \sigma'} c_{i \sigma}\right]
\end{multline}
We also demand that non-zero elements must have all four indexes $i$, $j$, $k$, $l$  distinct. Up to these symmetries, the matrix elements $J_{ij;kl}$ are assumed to be real independent random variables, drawn from the Gaussian distribution with the following mean and variance, 
\begin{equation}
    \langle J_{ij;kl}\rangle=0,\quad \langle J_{ij;kl}^2 \rangle=\frac{J^2}{(4N)^3}.
\end{equation}
It is supplemented with the attractive Hubbard interaction
\begin{multline}\label{eq:Hubbard}
    H_{\text{Hub}} = - U \sum_{i}^N c^{\dagger}_{i \uparrow} c_{i \downarrow}^{\dagger} c_{i \downarrow} c_{i \uparrow} - \mu \sum_{i,\sigma}^N c^{\dagger}_{i \sigma} c_{i \sigma} = \\ = - U \sum_{i}^N b^{\dagger}_i b_i - \mu \sum_{i,\sigma}^N c^{\dagger}_{i \sigma} c_{i \sigma},\quad b_i^{\dagger} = c^{\dagger}_{i \uparrow}c^{\dagger}_{i \downarrow}
\end{multline}
Hamiltonians above conserve particle number and are symmetric under the time-reversal transformations. States of these models are governed by temperature $T$, fermion occupation number $N_f$, along with the dimensionless parameter, $U/J$, characterizing the attraction strength.

In the absence of the SYK term the ground state of the pure Hubbard model, eq.~\eqref{eq:Hubbard}, consists of localized pairs and does not exhibit ODLRO. Its energy is obviously $-U$ per fermion pair and its degeneracy is given by the number of combinatorial possibilities of distributing a given number of pairs among $N$ orbitals. Excited states are formed by breaking some of the pairs and creating single occupied orbitals with zero energy. 

We shall be interested in the regime when the Cooper pairs get formed due to the Hubbard interaction however they are not necessarily get synchronized hence we have the system whose phase space is represented by the pairs $(\theta_i, p_i)$ where $\theta_i$ is the phase of the condensate of $i$-th Cooper pair and $p_i$ is the corresponding momentum.
The Hubbard interaction induces the all-to-all cosine interaction potential \cite{Gorsky2020} yielding the total Hamiltonian
\begin{multline}\label{eq:HMF-classical}
     H_{\text{SYK+U}}(\theta_i)=H_{\text{HMF}}(\theta_i)=\\=\sum_{i=1}^{N}\frac{p_i^2}{2m}-\frac{g}{N}\sum_{i<j}^{N}\cos\left(\theta_j-\theta_i\right).
\end{multline}
where $g$ is the coupling constant that has quantum nature \cite{Gorsky2020},
\begin{equation}
    g\sim C\frac{\Delta^2}{J} \propto J\exp\left(-\frac{\pi J}{U}\right)>0, \quad C\sim O(1)
\end{equation}
and $m$ corresponds to the susceptibility of ground state energy $E_{\text{GS}}$ to a local chemical potential $\mu$ in $N\rightarrow\infty$ limit \cite{gu2020notes}
\begin{equation}
    m\sim\frac{\partial^2 E_{\text{GS}}}{\partial\mu^2}\sim \frac{1}{J}
\end{equation}
In the article \cite{Gorsky2020} the quantum phase transition (QPT) has been analysed in mean-field approximation. It occurs for $g>g_c$ with $g_c=(2m)^{-1}$ and describes the transition from the pseudogap phase to the superconducting phase. Due to the all-to-all couplings (infinite range interactions) mean-field consideration gives an exact answer in $N\rightarrow\infty$ limit. In terms of soft-modes $\theta_i$, in the pseudogap phase we deal with non-synchronized state, whereas in the superconducting phase we have synchronized one. Using the relation with quantum HMF model at large $N$ limit we shall observe partially synchronized state in pseudogap phase as well. As we will describe below, it is caused due to the fact that classical QSS survive on a quantum level and therefore partial phase coherence can occur.

\section{Quantum HMF model and out-of equilibrium phenomena}

The equilibrium properties of the classical HMF model (\ref{eq:HMF-classical}) are comprehensively discussed \cite{Antoni1995} and presence of LRI is crucial, because it simplifies all the computations starting from the partition function to statistical averages. For non-equilibrium case, the thermodynamic limit, i.e. $N\rightarrow\infty$ can be obtained in the rigorous way with help of Braun-Hepp \& Neunzert theorems and yields the Vlasov equation \cite{Braun1977}. Roughly speaking, these theorems guarantee that trajectories in phase space obtained via Vlasov equation coincide with the solutions to equations of motion for classical HMF model in the limit of $N\rightarrow\infty$ \cite{Yamaguchi2004}. The appearance of Vlasov equation immediately reveals questions concerning a relaxation to Boltzmann equilibrium and classification of stationary solutions. Vlasov equation possesses an infinite set of conserved quantities --- Casimirs \cite{Bouchet2010,Elskens2014}. Presence of these quantities prevents relaxation to a Boltzmann equilibrium and causes stirring phenomenon in a phase space, which was firstly noticed by Lynden-Bell for gravitational systems \cite{Lynden1967}. This stirring (or mixing) in phase space can be interpreted as existence of quasi-stationary states (QSS). Dynamics of Vlasov equation favors formation of QSS and it was shown that their lifetime is very large \cite{Yamaguchi2004,Koyama2008}. The formation of QSS for classical HMF model is intensively studied for years and many interesting results were obtained. Among them, we would like to highlight cluster and bicluster QSS for attractive and repulsive HMF model \cite{Barre2002}, small travelling cluster states \cite{Barre2009}, and bicluster state for attractive model \cite{Antoniazzi2007}. Existence of these states implies that HMF model has a very rich out-of-equilibrium dynamics.

Inspiring by all these phenomena in the classical model, it is natural to ask for an impact of quantum effects for equilibrium and out-of-equilibrium properties of quantum HMF model which is necessary for interpretation in 
dynamics of phase modes in SYK+U model. Quantum HMF model has the following Hamiltonian
\begin{equation}\label{eq:HMF-quantum}
    H_{\text{HMF}} = -\frac{1}{2m}\sum_{i=1}^{N}\left(\frac{\partial}{\partial\theta_i}\right)^2-\frac{g}{N}\sum_{i<j}^{N}\cos\left(\theta_j-\theta_i\right),
\end{equation}
which describes $N$ interacting particles with mass $m$ on the unit circle. In case of $g>0$, we deal with attractive interaction, whereas in case of $g<0$ we obtain repulsive interaction. Following \cite{Chavanis2011Bosons}, we introduce rescaled Planck constant $\chi=1/\sqrt{m|g|}$ and rescaled time, $\tau=t/\sqrt{m|g|}$. Restoring Planck constant dimensionality and counting powers of $\hbar$, it is straightforward to see that $\chi\propto\hbar$, therefore $\chi\rightarrow 0$ corresponds to the classical limit. The classical equilibrium HMF model with $g>0$ exhibits the continuous phase transition with the order parameter,
\begin{equation}
    M=\frac{1}{N}\sum_{j=1}^{N}e^{i\theta_j},
\end{equation}
which is similar to magnetization in spin models. This continuous phase transition corresponds to the transition between the non-magnetized phase with $|M|=0$ and the magnetized phase with $|M|\neq 0$. In the thermodynamic limit $N\rightarrow\infty$ the quantum model is governed by generalized Gross-Pitaevskii equation (GGPE) \cite{Plestid2018},
\begin{gather}\label{eq:GGPE-HMF}
    i\chi\frac{\partial\Psi}{\partial\tau}=-\frac{\chi^2}{2}\frac{\partial^2\Psi}{\partial\theta^2}-\sgn(g)\Phi(\theta,\tau)\Psi,\\ \Phi(\theta,\tau)=\int_{-\pi}^{+\pi}d\theta'\cos\left(\theta-\theta'\right)|\Psi(\theta,\tau)|^2,
\end{gather}
where $\Psi$ is the condensate wave-function and the order parameter becomes
\begin{equation}
    M(\theta)=\int_{-\pi}^{+\pi}d\theta\,\cos(\theta-\phi)\rho(\theta,\tau),
\end{equation}
and $\phi=\phi(\tau)$ corresponds to the direction of the magnetization. The GGPE~\eqref{eq:GGPE-HMF} can be rewritten as quantum Euler equations with representing $\Psi=\sqrt{\rho}e^{iS/\chi}$ and $v=\partial S/\partial\theta$,
\begin{gather}\label{eq:Madelung-HMF}
    \frac{\partial\rho}{\partial\tau}+\frac{\partial}{\partial\theta}\left(\rho v\right)=0,\\ \frac{\partial v}{\partial\tau}+v\frac{\partial v}{\partial\theta}-\frac{\partial\Phi}{\partial\theta}=-\frac{\partial Q}{\partial\theta},\,Q\equiv\frac{\chi^2}{2\sqrt{\rho}}\frac{\partial^2\sqrt{\rho}}{\partial\theta^2}.
\end{gather}
The quantity $Q$ is the so-called quantum potential, initially introduced by Madelung and then used by Bohm in his formulation of quantum mechanics \cite{bohm1952suggested,wyatt2005quantum}. The thermodynamic limit of classical HMF model can be obtained by setting $Q=0$ (cf. with classical equations in \cite{Barre2002}). The first investigation of how the presence of quantum potential $Q$ affects the classical QSS have been performed by Chavanis \cite{Chavanis2011Bosons} and later examined by Plestid with coauthors in \cite{Plestid2018}. For the attractive interaction, the linear stability analysis of homogeneous solution, $\rho_0=(2\pi)^{-1}$, gives the following equation for perturbation $\delta\rho=\delta\rho(\theta,\tau)$, 
\begin{multline}\label{eq:Linear-Stab}
    \frac{\partial^2\delta\rho}{\partial\tau^2}=-\rho_0\frac{\partial^2\delta\Phi}{\partial\theta^2}-\frac{\chi^2}{4}\frac{\partial^4\delta\rho}{\partial\theta^4},\\ \delta\Phi=\int_{-\pi}^{+\pi}d\theta'\,\cos\left(\theta-\theta'\right)\delta\rho(\theta',\tau),
\end{multline}
and then representing perturbation as
\begin{equation*}
    \delta\rho(\theta,\tau)=\sum_k\delta\rho_k\exp\left(i\omega_k\tau-ik\theta\right),
\end{equation*}
we found that homogeneous solution $\rho_0$ is stable if $\chi>\chi_c=\sqrt{2}$. As should be, this result coincides with $g_c=(2m)^{-1}$ (in terms of $g$, the homogeneous solution $\rho_0$ is stable for $g<g_c$). It means that QPT with respect to coupling constant $g$ takes place, which raises the formation of spatially inhomogeneous condensate of soft modes, i.e. synchronized state with $|M|\neq 0$. The explicit expression for inhomegenous solution can be obtained by self-consistent solving of the Mathieu equation \cite{Plestid2019}. Careful consideration shows that there is a tower of condensate characterized by the number of condensate wave-function nodes, where each wave-function corresponds to the solution of Mathieu equation.

\begin{figure}
    \centering
    \includegraphics[width=0.8\linewidth]{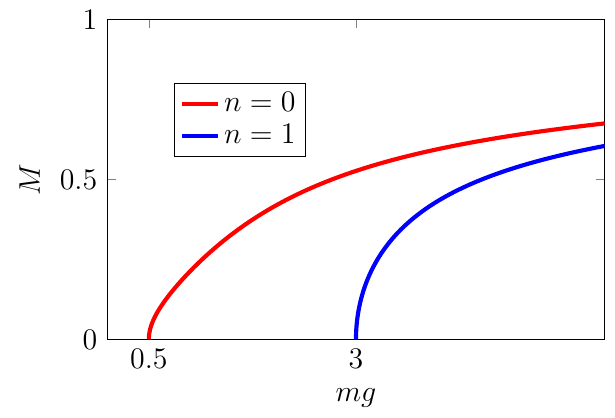}
    \label{fig:phase_diagram}
    \caption{Equilibrium phase diagram for the SYK+U model in the $U/J\ll 1$ limit obtained by self-consistency solutions to the Mathieu equation ($n$ denotes the number of nodes of condensate wave-function)}
\end{figure}
\begin{figure}
    \centering
    \includegraphics[width=\linewidth]{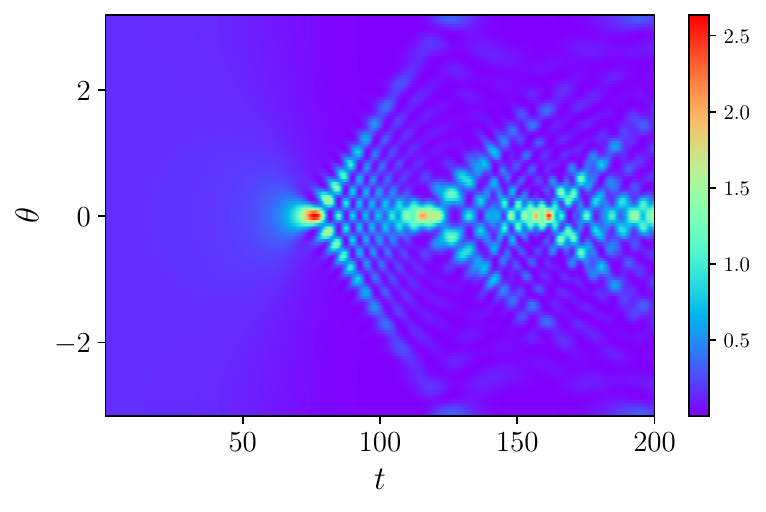}
    \label{fig:quantum_qss}
    \caption{Density plot of $\rho=\rho(\tau,\theta)$ corresponding to quantum QSS for HMF model with $\chi=0.1$}
\end{figure}

The presence of quantum potential is crucial because it stabilizes the homogeneous solution. It was shown that in case of small but non-zero values of $\chi$ for attractive interaction the QSS appears. This QSS is well-known and it is the quantum twin of single cluster QSS in classical HMF model (see fig.~\ref{fig:quantum_qss}). Note that this state is well defined in large-$N$ limit. The case of repulsive classical HMF model is even more intriguing. It was shown that in such case in thermodynamic limit the bicluster QSS appears \cite{Barre2002}. The existence of bicluster is caused by the presence of two drastically different time-scales, an analog of plasma frequency (large time scale) and energy per particle (fast time scale). The bicluster persists in quantum repulsive HMF model for small values of $\chi$, which also was shown in \cite{Plestid2018}.

We can also treat the quantum HMF model in terms of Wigner equation, which plays role of quantum Vlasov equation \cite{Chavanis2011Bosons}. This approach is important because GGPE corresponds to the zero-temperature case, whereas using quantum Vlasov equation, it seems possible to analyze finite temperature case as well. We start from the quantum Liouville equation for the density matrix and construct the Wigner function ,
\begin{multline}
    f(\theta,\,p,\,\tau)=\\=\int_{-\pi}^{+\pi}\frac{d\phi}{2\pi\chi}e^{-ip\phi/\chi}\rho\left(\theta+\frac{\phi}{2},\,\theta-\frac{\phi}{2},\,\tau\right).
\end{multline}
where $\rho$ is the density matrix. After straightforward manipulations, we write down the Wigner equation for the quantum HMF model,
\begin{multline}
    i\chi\frac{\partial f(\theta,\,p,\,\tau)}{\partial\tau}=-i\chi p\frac{\partial f(\theta,\,p,\,\tau)}{\partial\theta}+\\+\left\{\int_{-\pi}^{+\pi}d\alpha\int_{-\infty}^{+\infty}dk f(\alpha,\,k,\,\tau)\left[\cos\left(\theta-\alpha-\frac{i\chi}{2}\frac{\partial}{\partial p}\right)\right.\right.-\\-\left.\left.\cos\left(\theta-\alpha+\frac{i\chi}{2}\frac{\partial}{\partial p}\right)\right]\right\}f(\theta,\,p,\,\tau).
\end{multline}
where $f=f(\theta,\,p,\,\tau)$ is the Wigner function and the equation is written down for the attractive model. Linearization around the spatially homogeneous solution $f_0(p)$ gives the well-known result for stability of homogeneous solutions, $\chi>\chi_c$. As should be, the classical Vlasov equation can be obtained by taking $\chi\rightarrow 0$ limit. It is worth mentioning that quantum Vlasov equation approach creates an opportunity to investigate more complicated dynamics. Also, note that we can relate GGPE to the quantum Vlasov equation integrating the Wigner function over the momenta.

Summarizing we can predict the following phenomena in the pseudogap phase of SYK+U using its relation with HMF model:

\textbf{Quantum QSS states.} Going to pseudogap phase, we prepare the initial state that is characterized by phases $\theta_i(t=0)$ and time derivatives of phases $\dot{\theta}_i(t=0)$. Then we analyze how this quench evolves in time in accordance with HMF model dynamics and this setup coincides with investigation of quench relaxation in BCS model. During the evolution, the interference of states in Fock space occurs \cite{Mumford2019} and for small enough quantum effects QSS appears, which causes the superconductivity fingerprints in pseudogap phase. The fact that QSS is still alive in quantum HMF model allows us to predict that more exotic states have a right to exist. This proposition raises a discussion about an interconnection between stirring phenomenon in phase space and quantum states interference in Fock space.

\textbf{Phase transitions in dependence of initial conditions.} The typical parameter space of classical HMF model is covered by the coupling constant $g$ and the initial magnetization $M_0$. It is easy to show that the value of coupling constant reflects in the internal energy density, i.e. energy per particle $E$. For instance, in case of so-called waterbag initial conditions, where the phases are picked randomly from the interval $[-\theta_0,+\theta_0]$ and momenta are also randomly picked from the interval $[-p_0,+p_0]$, one can show that energy per particle and initial magnetization are determined by
\begin{equation*}
    M_0=\frac{\sin(\theta_0)}{\theta_0},\quad E=\frac{p_0^2}{6}+\frac{1-M_0^2}{2}.
\end{equation*}
Looking for the stationary solutions of the corresponding Vlasov equation, one can see the line in $(E,M_0)$-plane that separates the solutions with zero and non-zero magnetization. This transition is the first order phase transition. Comparing with SYK+U, we conclude that in $U/J\ll 1$ limit, the similar phase transition should take place. However, in order to identify such transition, one should carefully investigate the Wigner equation, which we postpone for further research.
    
\textbf{Higher condensates.} For the equilibrium properties, there is one more interesting feature for the quantum HMF model. Careful treatment of equilibrium solutions of corresponding GGPE was done in \cite{Plestid2019}. This consideration demonstrates that in quantum HMF there is the tower of condensates, characterized by different Mathieu functions. These condensates appear as the solitons of Mathieu equation, where the amplitude of interaction is computed in self-consistent way. This means that in the SYK+U model we have tower of condensates in the superconducting phase (see fig.~\ref{fig:phase_diagram}).

As we have mentioned earlier, the stirring in phase space can be probed in Wigner function framework, which corresponds to the semi-classical limit. It was pointed out in \cite{Barre2002}, the emergence of bicluster is related to so-called chevrons, that correspond to singularities in phase space. Such singularities are treated by a catastrophe theory, which uses ADE-classification for the patterns of  singularities. On the quantum level, ADE-classification still takes place and each catastrophe is characterized by its own generating function, that was reviewed in great details in \cite{Kirkby2022}. For the quantum HMF model, we deal with cusp catastrophe, which is generated by Pearcey function \cite{Plestid2018}. 

\section{Comment on the holographic picture}

Let us make a few short comments concerning the possible counterparts of the phenomena discussed above in the holographic dual postponing more detailed analysis for the separate study. The complex SYK model itself at low temperature is dual to the JT gravity supplemented with the Abelian gauge field \cite{gaikwad2020holographic,chaturvedi2021ads3}. This dual reproduces correctly the effective action for the phase field \cite{davison2017thermoelectric}. We however look for the transition to the superconducting phase which implies the necessity to add some ingredients providing the Cooper pair condensate formation in the dual picture. The quite quantitative derivation of the holographic dual has been found in \cite{inkof2022quantum} for the SYK+Yukawa model. The dual picture involves the AdS$_2$ geometry, gauge field and massive complex scalar. In this approach the holographic picture matches the Eliashberg equations and the negative mass of the scalar indicates the instability of the normal state. However this picture is not suitable for the discussion of the pseudogap phase.

To get the qualitative holographic explanation of out-of-equilibrium phenomena in the pseudogap phase we follow another approach. The low energy effective action for pure  complex SYK model can be derived from the AdS$_3$ gravity via the KK reduction \cite{gaikwad2020holographic,chaturvedi2021ads3}. One starts with the BTZ black hole solution to 3D gravity which involves two $S^1$ circles with coordinates $(\theta,\tau)$, where $\tau$ is Euclidean time coordinate and supplement the 3D action with the $\mathrm{U}(1)$ Chern-Simons term. It was argued in \cite{gaikwad2020holographic} that KK reduction of the action with respect to the non-contractable $\theta$-circle amounts to the Schwartzian action for the pseudo-Goldstone mode  $f(\tau)$ . The kinetic terms for the soft mode for $\mathrm{U}(1)$ gauge field \cite{davison2017thermoelectric,gu2020notes} follows from the Chern-Simons term.

To describe the transition to superconducting state we add the complex scalar which indicates the condensate formation $\Psi(r,\theta,\tau)=|\Psi| \exp(i\phi (r,\theta,\tau))$ to the 3D BTZ geometry supplemented with CS term. The kinetic term for the scalar involves the covariant derivative $D_{\mu}=\partial_{\mu}-2iqA_{\mu}$ for the field with charge $2q$. The Hubbard interaction induces the Cooper pairing and therefore yields non-vanishing modulus of the scalar $|\Psi|$. Not much is known about the holographic counterpart of the Hubbard coupling and we restrict ourselves by interpretation suggested in \cite{fujita2019effective} where the Hubbard coupling provides a kind of cut-off in the radial coordinate. Recall that holographically isolated fermion is the string extended along the radial coordinate. Hence we have distribution of the string ends along the $\theta$-circle. The presence of CS term indicates that we have the flavor D-brane extended along this circle therefore the open strings can end on the brane. Hence we can naturally introduce the collective phase field $\phi(r,\theta,\tau)$ in the bulk. 

The focus in our study concerns the inhomogeneous distribution of the phases of the Cooper pairs in the pseudogap phase which gets promoted to the $(\theta,\tau)$ dependent distribution in the thermodynamic limit of the HMF model. We conjecture that this distribution or equivalently collective field in the boundary theory has the $\phi(r,\theta,\tau)$ field as the bulk dual. We assume that the holographic non-contractable $\theta$ coordinate hosts the phases of individual Cooper pairs $\theta_i$. When there is no synchronization of phases of Cooper pairs $\theta_i$ the phase of scalar field depends on $\theta$ non-trivially. In this case it is impossible to perform the KK reduction in the simple way and the dual theory remains three-dimensional. Somewhat similarly to holographic QCD in another gauge the following representation of the scalar is possible
\begin{equation}
    \phi(\theta,\tau)= \int dr A_r(r,\theta,\tau)
\end{equation}
in terms of radial holonomy of the gauge field. To justify this picture quantitatively it is necessary to identify the holographic meaning of the Hubbard term more carefully.

The BTZ geometry involves the thermal circle therefore we effectively 
work with the Euclidean time in the boundary theory. Since in the real time the HMF Hamiltonian provides the attraction and synchronization in the Euclidean time we have the repulsive interaction in the HMF Hamiltonian and the corresponding out-of-equilibrium phenomena. In particular we predict in the holographic dual assuming the holographic duality with HMF model

\textbf{QSS states in the 3D dual system.} At the classical level boundary HMF model for the repulsive interaction enjoys the long-lived bicluster state for a wide class of the initial conditions. It can be recognized from the characteristics of the forced Burgers equation derived from the Vlasov equation at small temperature limit
\begin{equation}
    \frac{\partial v}{\partial \tau } + v\frac{\partial v}{\partial \theta } = - \frac{1}{2}\sin 2\theta
\end{equation}
which at early time admits the shock wave and caustics with some periodicity. We expect the similar QSS state for the scalar field around the BTZ geometry and the Vlasov equation presumably can be derived from the 3D theory supplemented with the Hubbard perturbation which yields the sine term.

Note that presumably more general gauge group in the bulk could be considered. According to the holographic duality the global symmetry at the boundary gets promoted to the gauge theory in the bulk. If we assume global symmetry
rotating each fermion flavor independently with additional constraints imposed by interaction terms  the boundary flavor group gets mapped to more general gauge group in the bulk somewhat similar to the holographic QCD.

\section{Discussion}

In the superconductivity language, the appeared phenomenon is tightly connected to the dynamical effects known for BCS model \cite{Yuzbashyan2005,Yuzbashyan2006,Barankov2006} and to so-called Higgs mode engineering \cite{shimano2020higgs}. Despite appealing similarity, in both cases one deals with time-dependent superconducting gap, whereas in SYK+U model we deal with the time-dependent phase of gap and in this sense it is quite different phenomenon. The most significant step during the analysis of BCS model dynamics is to rewrite of BCS Hamiltonian in terms of pseudospins,
\begin{equation}
    H_{\text{BCS}}=\sum_{j=0}^{N-1}2\epsilon_jK_j^z-g\sum_{j,q}K_j^{+}K_q^{-},
\end{equation}
where $g$ is the BCS coupling constant, $K_j^{-}=c_{j\uparrow}c_{j\downarrow}=(K_{j}^{+})^{\dagger}$, $K_j^{z}=(n_{j\uparrow}+n_{j\downarrow}-1/2)$, $\epsilon_j$ represents the energy of $j$-th orbital, $n_{j\sigma}=c^{\dagger}_{j\sigma}c_{j\sigma}$ and $N$ is the number of particles. The resulting Hamiltonian raises the Bloch form dynamics. In this set up one also has LRI due to the all-to-all couplings. It means that in case of large $N$ limit, $N\rightarrow\infty$, the mean-field description is exact. Moreover, this model is directly connected to so-called central spin model, where the central spin interacts with environmental spins. It is interesting that both these models are integrable and their integrability was analyzed comprehensively \cite{Yuzbashyan2005solution}. It was shown in \cite{Barankov2006} that emergent dynamical effects in the BCS model are related to linear Landau damping with generalizations for non-linear case. It is important  since such effects arise as interplay between collective mode and continuous part of spectra. In case of quantum and classical HMF model, we also deal with interaction between collective mode (wave) and particles, which is common situation in plasma. However, here the similarity between these two models is over.

The appearance of cluster \& bicluster QSS in quantum HMF model share similarities with the concept of quantum revivals \& quantum many-body scars. In the classical HMF model, the time-period of bicluster can be found explicitly with help of analysis of forced Burgers equation \cite{Barre2002}. In case of cluster QSS, for our best knowledge there is no estimation of cluster appearance period. In the recent paper by Defenu \cite{defenu2021metastability}, the author demonstrates that the energy spectrum of long-range interacting systems shares some similarity with system with disorder. It is directly related to the behavior of recurrence time in $N\rightarrow \infty$ limit. In addition, recently the discussion of the quantum revivals was continued in a different context in \cite{Alhambra2020}, where the authors illustrates how the revivals in the model system are related to quantum many-body scars.

These eye-catching results imply the possible relations  between quantum caustics and quantum many-body scars. To pursue this idea one has to carefully consider the dynamics of Vlasov equation which causes the filamentation in phase space, which corresponds to the formation of thin filaments in phase space from the initial state. To form an idea of a quantum analog of this phenomenon, one can think about Wigner function (at this point, we would like to draw attention for the discussion of phase space formulation of quantum dynamics, \cite{polkovnikov2010phase}). At the microscopic level, this process continues for all times. The Vlasov distribution function $f$ conserves so-called Casimirs, for instance moments of $f$,
\begin{equation}
    \mathcal{C}_n=\int d\theta\,dp\,f^n
\end{equation}
More generally, for any convex function $F$, the following quantity
\begin{equation}
    \mathcal{C} = \int d\theta\,dp\,F[f]
\end{equation}
is conserved under the Vlasov dynamics. However, as was pointed out in \cite{Levin2014}, at the coarse-grained level, the evolution ends and the asymptotic distribution function $\overline{f}$ appears and one can show that it can be described in terms of the Lyndenn-Bell distribution. The resulting distribution function feels the initial conditions and the moments of coarse-grained distribution are not conserved (in compare with fine-grained one) \cite{chavanis2022kinetic}. The properties of such distributions, in particular, stability analysis is intensively studied from the early works \cite{Yamaguchi2004}. In addition to described above properties of the Vlasov equation, we would like to emphasize that systems with LRI admit weak ergodicity breaking, which was discussed in \cite{Benetti2012,Figueiredo2014}. 

Considering all the above, we map these statements onto our case and propose that the SYK+U model in the pseudogap phase exhibits weak ergodicity breaking too and the underlying mechanism is related to the emergent integrability of the Vlasov equation.

\section*{Acknowledgements}

A.A. is grateful to G.V. Zasko for valuable discussion of numerical solution of GGPE equation and to D. O'Dell for comments concerning the simulations of QSS in quantum HMF model. A.G. thanks Nordita and IHES for the hospitality and support. The work of A.A. was supported by the Foundation for the Advancement of Theoretical Physics and Mathematics ``BASIS'' (grant №23-1-5-41-1) \& by Brain Program of the IDEAS Research Center.

\bibliography{references.bib}

\end{document}